\documentclass[conference]{IEEEtran}
\usepackage{listings}
\usepackage{graphicx}
\usepackage[english]{babel}
\usepackage{hyperref}
\newcommand{\starss}{{StarSs}}
\newcommand{\smpss}{{SMPSs}}
\newcommand{\lbc}{{\em lbc}}
\newcommand{\laki}{{\em Laki}}
\begin{document} 

\title{Hybrid MPI/\starss{} -- a case study}
\author{
  \IEEEauthorblockN{
    Jos\'e Gracia\footnote{gracia@hlrs.de}\IEEEauthorrefmark{1}, 
    Christoph Niethammer\IEEEauthorrefmark{1},
    Manuel Hasert\IEEEauthorrefmark{2},
    Steffen Brinkmann\IEEEauthorrefmark{1},
    Rainer Keller\IEEEauthorrefmark{1},
    Colin W. Glass\IEEEauthorrefmark{1}}
  \IEEEauthorblockA{\IEEEauthorrefmark{1}High Performance Computing
    Center Stuttgart (HLRS), University of Stuttgart, 70550
    Stuttgart, Germany}
\IEEEauthorblockA{\IEEEauthorrefmark{2}German Research School for Simulation Sciences
   GmbH, 52062 Aachen, Germany}
}
\maketitle
\begin{abstract}
  Hybrid parallel programming models combining distributed and shared
  memory paradigms are well established in high-performance
  computing. The classical prototype of hybrid programming in HPC is
  MPI/OpenMP, but many other combinations are being
  investigated. Recently, the data-dependency driven, task parallel model
  for shared memory parallelisation named \starss{} has been suggested
  for usage in combination with MPI. In this paper we apply hybrid
  MPI/\starss{} to a Lattice-Boltzmann code. In particular, we present
  the hybrid programming model, the benefits we expect, the
  challenges in porting, and finally a comparison of the
  performance of MPI/\starss{} hybrid, MPI/OpenMP hybrid
  and the original MPI-only versions of the same code.
\end{abstract}

\section{Introduction}

The Message Passing Interface (MPI) is the \emph{de facto} standard in
high performance computing. While MPI is a distributed memory
parallelisation scheme, it is also frequently used on SMP-like
systems, as for instance today's multi-core CPUs. On such systems,
the parallel applications developer may resort to shared
memory parallelisation schemes as for instance OpenMP. Hybrid
parallelisation models consisting of a distributed memory part to
pass messages between compute nodes in a cluster, and a
shared memory part to exploit all available cores on a node, have been
used in HPC with success.
Yet, the general perception is, that MPI-only jobs
are still the working horse in research groups relying on numerical
computations. 

The MPI-only approach has come under pressure; mainly it is challenged by today's prevalence of
multi-core and many-core systems. One aspect of the problem
certainly is that many applications are structured in such a way that
they cannot scale efficiently to large numbers of MPI
ranks; they have limits on MPI-scalability that may well
depend on problem size or simulation parameters. Often these limits
could be lifted or pushed further out by an algorithmic
redesign. However, most research groups will \emph{not} go down such a road
lightly.

In this paper we will not delve into the lack of scalability of MPI
itself, whether perceived or real (see \cite{B+09} for a discussion on
  the topic), but will assume that an
application's limit on MPI-scalability is inherently a problem of a
given application itself. Users hitting the limits of such an MPI-only
application have three choices: 1) find a set of parameters
(e.g. problem size) that allows to scale out further, 2) redesign and
rewrite their application to use MPI more efficiently, 3) go hybrid,
which also means redesigning and rewriting the code. Often the first
is not possible or interesting from a scientific point of view.  From
the last two choices, the hybrid approach is usually more straightforward, as
the re-design can be implemented incrementally, allowing to
assess its correctness along the way. Note, however, that in view of
Amdahl's law, any hybridization needs to cover the code in its
completeness in order to be efficient.

The standard paradigm for hybrid parallel programming in HPC is MPI/OpenMP,
but many other combinations are being investigated. Recently, the data-flow
driven, task parallel model for shared memory parallelisation named
\emph{\starss{}} has been suggested as a partner for MPI. In this paper, we
will first introduce the main aspects of \starss{} briefly. Then we will
apply the hybrid MPI/\starss{} model to an MPI-only Lattice-Boltzmann code
while describing the challenges we encountered and how they have been solved.
Furthermore, a hybrid MPI/OpenMP version is developed and used as a basis 
for the assessment of the hybrid MPI/\starss{} version.
Next, we show performance of the current hybrid implementation, contrast it
to the original, and finally draw our conclusions.

\section{The hybrid MPI/\starss{} programming model}

We start by giving a short overview of the \starss{} programming
model~\cite{*Ss:BPBL06}. It is based on task parallelisation.  First
of all, the programmer needs to identify suitable units of work, in
general at subroutine boundaries, and designate them as tasks through
pragma based code annotations. 
In contrast to, e.g., pthreads or OpenMP, where task synchronization
has to be specified by the programmer explicitly, \starss{} infers the synchronization from
data dependencies in the program. These dependencies are automatically
detected, based on the directionality of subroutine arguments,
using code directives to distinguish between
input, output and input-output arguments. If programming in Fortran
the already present 
\verb!intent(...)!  clauses in subroutine interfaces are used.

At runtime, the data directionalities and the actual data arguments,
i.e. their memory address, are taken to build up a dynamic task
dependency graph. In the simplest case, tasks will be scheduled for
execution sequentially in the order in which they have been added to
the graph. This is essentially equal to serial execution. In general, the
task graph will expose concurrency which can be exploited by
dispatching independent tasks onto a range of available compute cores and running these
in parallel. Data dependencies then will ensure that no task is
scheduled before all task that modify its inputs have completed. 
The \starss{} model allows to write programs without any kind of
explicit synchronisation; in fact, this is the rule. Synchronisation
is in principle necessary only when code executed synchronously
outside of tasks (executed by a master thread) and code executed
inside of tasks (executed asynchronously by worker threads) use the same
data. In this sense synchronisation is required only between
sequential code running on the master thread and tasks, not between
tasks. Naturally, one can avoid this kind of master-worker
synchronization by taskifiying all relevant parts of the code.

Compared to classical parallelisation approaches using parallel loops
or explicitly synchronised tasks, \starss{} has several
advantages. Firstly, it is a dynamic process which will allow
efficient parallelisation even for different input sets; each run
might result in a different dependency graph and thus in different
program execution behaviours. Secondly, and much more importantly, the
lack of explicit synchronisation between tasks can improve load
balancing and solve some of the scalability problems.

The third point to be mentioned concerns the hybrid MPI/\starss{}
approach; MPI communication can also be off-loaded into \starss{}
tasks. This allows overlapping of computation and communication in a
simple and more efficient way than hand-written code using,
e.g.,  non-blocking MPI operations. In this paper we will explore this
possibility only briefly, however.

In some sense, \starss\ is a family of programming models including,
among others, SMPSs and OmpSs. All of them very similar in spirit,
particularly the focus on data-dependencies between tasks, but with
slight differences in the specifics of the compiler directives and the
capabilities of the runtime. For this work, we have used the SMPSs
compiler, which is tailored to multi-core and SMP-like machines, but
also aware of MPI allowing to off-load MPI tasks onto a dedicated
communication thread. We did not consider the more recent alternative
OmpSs, simply because it did not support Fortran at the time of
writing this paper.

\section{A case study: Lattice-Boltzmann}
Before starting to port an application to a new programming model,
most application developers in applied computational science will
balance expected performance gain against the expected effort. Any new
programming model should therefore not only promise high performance,
but should also be simple to use.  As presented in the previous section,
applying the \starss{} programming model to an existing application
is in principle straightforward as long as the data-flow is clearly
exposed in the given code. We have used \starss{} to
hybridise a previously MPI-only fluid dynamics code which implements
the Lattice-Boltzmann Method (LBM). In this paper we present our
parallelisation approach, discuss the challenges we encountered,
suggest how to overcome them, and finally put the resulting
performance into perspective.

We have purposely chosen to undertake this experiment with a code
written by a graduate student, which is not heavily optimised
regarding scalar or MPI performance. \lbc{}~\cite{HKR11} -- short for {\em
  l}attice-{\em b}oltzmann {\em c}ode -- is a Lattice-Boltzmann code
using the BGK~\cite{BGK54} approximation for the collision term. The
algorithm is a standard two-grid approach, i.e. at every algorithmic step 
the code reads from one grid and writes into the other, without any spatial
or temporal blocking. The code is written in Fortran90 and uses
modules, pointers, and user-defined types throughout. Similarly, the
MPI parallelisation is based on a simple multi-dimensional domain
decomposition. Ghost cells are exchanged at the end of each timestep,
no attempt to overlap computation and calculation is done.

The Lattice-Boltzmann method consist of three consecutive
computational steps -- \verb!stream()!, \verb!collide()!, and
\verb!boundaries()!. In order to increase data reuse, the former two
are often combined into a \verb!stream_collide()!, while the latter
applies external boundary conditions. In the distributed memory case,
ghost cells are exchanged with MPI neighbours in
\verb!exchange_ghosts()!.  In our case the main time-stepping loop is
\begin{lstlisting}
do timestep=1, timesteps
  call stream_collide(block)
  call boundaries(block)
  call exchange_ghosts(comm, block)
end do
\end{lstlisting}
where \verb!block! is a variable of user-defined type \verb!type(lb)! encapsulating
all data structures necessary for the Lattice-Boltzmann algorithm,
most importantly the working arrays \verb!flip! and \verb!flop!;
\verb!comm! is another user-defined type encapsulating information
regarding MPI communication as for instance the rank of neighbours, etc.

\subsection{Tiling of the algorithm with data copies} \label{sec:copies}

\starss{} is a task parallel programming model. We want to
apply it to an algorithm, i.e. LBM, which is as many others in CFD
naturally data parallel, but at first glance does not seem to have more
than a few  obvious subroutines suitable as tasks. Even
worse, the candidates \verb!stream()!, \verb!collide()!,
\verb!boundaries()!, and \verb!exchange_ghosts()! have flow
dependencies on each other and may not be executed concurrently. 

In order to apply \starss{} to this code, we need to block (or tile) the
algorithm, in order to convert data parallelism to task
parallelism. In the following we use the term \emph{tiling}.
Rather than tiling on the loop level, we do so at the
top program level by introducing an additional domain decomposition
for tiling on top of the MPI domain decomposition. In the
following we will refer to an MPI subdomain as {\em block} and to a
\starss{} subdomain as {\em tile}. Note, that tiles are subdomains of a
particular block.  In the code, the variable \verb!block! represents
a block and tiles are represented by \verb!tile!. For simplicity
each tile is based on the same user type as blocks allowing
to reuse much of the original code. 

Similar to blocks on the MPI level, tiles need to exchange ghost
cells with every neighbour. So, before doing the LBM step, tiles pull
ghost cells from a buffer (\verb!pull_ghosts()!) and push them back into a
buffer (\verb!push_ghosts()!) after concluding the LBM step. We have chosen to use the MPI
level block as buffer. The tiled main time-stepping loop is then:
\begin{lstlisting}
call init_tiles(block, tiles)

do timestep=1, timesteps
  do iTile=1, nTiles
    tile => tiles(iTile)
    call pull_ghosts(tile, block)
  end do

  do iTile=1, nTiles
    tile => tiles(iTile)
    call stream_collide(tile)
    call push_ghosts(tile, block)
  end do

  call boundaries(block)
  call exchange_ghosts(comm, block)
end do

call gather_tiles(block, tiles)
\end{lstlisting}
 Note, that inside the time-stepping loop \verb!block! only carries
 valid data for those array elements that are required to exchange ghost
 cells, either with MPI neighbours in \verb!exchange_ghosts()! or with
 neighbour tiles in \verb!push/pull_ghosts()!, respectively. At the end
 of the timestep loop, or before any output for that matter, data in the tiles
 needs to be gathered back into the block, which is done in
 \verb!gather_tiles()!. Similarly,  \verb!init_tiles()! initialises
 tiles from data in the block.

\subsection{Tiling of the algorithm with data
  aliasing} \label{sec:aliasing}

A major drawback of the tiled algorithm described in the section
above, is the data copies between the tiles and the block. This is true even if
only those array elements are copied which serve as ghost cells for adjacent
tiles or for neighbouring MPI blocks. The resulting code suffers
significant performance penalties in the order of several tens of
percent, as will be detailed later on.

Rather than replicating block's data in the tiles, we sought a way of
reusing the data in a block, by mapping or aliasing it into the tiles. 
We made use of the fact that in Fortran, array pointers can not only point to whole
arrays, but also into sub-arrays as
\begin{lstlisting}
  tile%flip => block%flip(.. ,xl:xu, ..)
  tile%flop => block%flop(.. ,xl:xu, ..)
\end{lstlisting}
where the array boundaries in the block, i.e. $xl$, $xu$, etc, are
distinct for each tile.  Note, that in general the pointers will alias
sub-arrays which are non-contiguous in memory. With this
implementation tiles and the block are just different views on the
same underlying memory region. The data is organised in such a way, that
it simplifies data access in different parts of the code, respectively.

Obviously, it is no longer necessary to keep tiles and blocks
consistent, nor to copy ghost cells between adjacent tiles. The
subroutines \verb!pull_ghosts()!, \verb!push_ghosts()!, and
\verb!gather_tiles()! are no longer needed. The only purpose of
\verb!init_tiles()! is to setup the aliases for \verb!flip! and
\verb!flop!. The basic algorithm thus simplifies considerably to
\begin{lstlisting}
call init_tiles(block, tiles)

do timestep=1, timesteps
  do iTile=1, nTiles
    tile => tiles(iTile)
    call stream_collide(tile, block)
  end do

  call boundaries(block)
  call exchange_ghosts(comm, block)
end do
\end{lstlisting}
Note, that this code resembles very closely the original non-tiled
version. The only difference apart from the initialisation of tiles is
the additional tiles loop embedding the call to the main computational
step \verb!stream_collide()!. The inclusion of the argument
\verb!block! to task \verb!stream_collide()! will become clear in
section \ref{sec:issues}.

\subsection{Task identification and data directionality}
\label{sec:tasks}

All subroutines in the main time-stepping loop are in principle suitable as
task. In \starss{} neither scheduling of tasks nor synchronisation amongst
them is specified explicitly. Instead, the developer has to declare the
directionality for each argument of any task subroutine, i.e. input, output,
or inout. In C, this is done through additional clauses to the \starss{} task
directive. The Fortran interface uses the Fortran clause \verb!intent(...)!,
which becomes mandatory for each argument of task subroutines. The interface
of our tasks is:

\begin{lstlisting}
interface
  !$CSS TASK REDUCTION(sentinel)
  subroutine stream_collide(tile, sentinel)
    type(lb), intent(inout)  :: tile
    type(lb), intent(inout)  :: sentinel
  end subroutine
  !$CSS TASK
  subroutine pull_ghosts(tile, block)
    type(lb), intent(inout) :: tile
    type(lb), intent(in)    :: block
  end subroutine
  !$CSS TASK REDUCTION(block)
  subroutine push_ghosts(tile, block)
    type(lb), intent(in)     :: tile
    type(lb), intent(inout)  :: block
  end subroutine
  !$CSS TASK
  subroutine boundaries(block)
    type(lb), intent(inout)  :: block
  end subroutine
  !$CSS TASK
  subroutine exchange_ghosts(block)
    type(lb), intent(inout)  :: block
  end subroutine
end interface
\end{lstlisting}

Note, that without the reduction clause,
i.e. \verb!REDUCTION(block)!,  on task \verb!push_ghosts()! the application
would be partly serialised: each invocation of \verb!push_ghosts()!
would depend on the previous one through the argument \verb!block!
with \verb!intent(inout)!. Thus, each instance of this task function
could not be executed before the previous one completed execution. The
reduction clause instructs the \starss{} runtime to ignore
inout-dependencies on a given datum for tasks participating in the
reduction, but retains dependencies to tasks not being part of the
reduction. It is then the responsibility of the developer to make sure
that parallel execution of reduction tasks does not lead to
data-races, e.g. by making sure that different tasks invocations modify only
non-overlapping data regions, or by using locks or atomic instructions.

It is worth noting that -- conceptually -- the most difficult part of
using \starss{} is constructing data dependencies in a way, that
allows sufficient concurrency to be identified and therefore exploited by the runtime. Another
difficulty is verifying that the directionalities specified by the
developer through \starss{} directives are compatible with the actual
behaviour of the code. In both cases the \starss{} debugger
\textsc{Temanejo}~\cite{BNGK11a,debugger:web} is a very
useful tool. It shows a visual representation of the actual dependency
graph at runtime and has the ability to control the scheduling of tasks.


\subsection{Fortran issues}
\label{sec:issues}

This study was done using the \smpss{} flavour of the \starss{}
programming family, more specifically \smpss{} V2.5 compiler
and runtime~\cite{*Ss:SMPSs25} were used. Unfortunately, this version still suffers
from a couple of problems regarding Fortran90 support. 

The most important issue is that subroutines declared in modules may not be
used as task functions, i.e. task functions are not allowed to be
part of a module. We have worked around this issue, by declaring a
wrapper subroutine outside of any module. The wrapper basically just
calls the original subroutine after \verb!USE!ing the respective
module.

The next issue is that the \smpss{} compiler and runtime system are
not able to track dependencies on a subset of data, in the sense of a
sub-array \verb!a(2:3)!  being a subset of the full array
\verb!a(:)!. Currently, \smpss{} has no notion of data {\em
  size} (at least when it comes to dependency-tracking). 
If one task has \verb!intent(out)! on \verb!a! and another
one has \verb!intent(in)! on \verb!a(2:3)!, the runtime will not
acknowledge a dependency between those two tasks, as it only
conceptualizes data based on its starting address in memory. A similar
situation arises in our case: each \starss{} tile is a subset of the
MPI block (even if there is no overlap in memory of the variables
\verb!tile! and \verb!block!), but there is no way to convey this
information to the runtime.

In our case, the Fortran modules were no real problem. Using wrappers
as workaround was simple.  A shortcoming resulting from this layer of
wrappers across Fortran modules is, that it prevents the
compilers from doing certain optimisations. The performance penalty in our
case was at the sub-percent level and thus acceptable.

It is difficult to give a general solution to the data subset
problem. One approach that has been suggested~\cite{*Ss:PBL08} is to
use {\em sentinels}, sometimes also called representants. One datum,
or rather its memory address, is chosen to represent a whole
collection of data. This is similar to colouring array elements in
loops. The sentinel is then passed as additional argument to tasks
subroutines with a corresponding intent. In principle the sentinel can
be any variable, in the simplest case an array with a single
element. Sometimes it is preferable to use an existing variable in
order to prevent clobbering the source code with additional variables without
apparent function. For instance, we have used \verb!block! as a
sentinel and passed it to tasks without actually using it inside the
task subroutines. This is the only reason, why our task
\verb!stream_collide()! takes \verb!block! as
argument.
Note, that a
reduction on \verb!block! is necessary to allow all instances of this
task to run in parallel; omitting this would result in a serialization
of all tasks. Note also, that in the version presented in
section~\ref{sec:copies} no sentinels are required as the task
\verb!push_ghost()! connects all tiles with the block dependency-wise.

\subsection{Performance measurements}
\label{sec:performance}

We have benchmarked three different versions of the applications. All
versions were compiled with GNU gfortran 4.6.1 as (backend)
compiler. The first is the original MPI-only version without any
modifications for \starss{} and was compiled with the Open~MPI 1.5.4
compiler driver. The second, \emph{MPI/\starss{} or hybrid version},
is the version described in this paper, i.e. has been refactored to
use tiles with aliasing (as describes in section \ref{sec:aliasing})
and works around the Fortran issues above. Throughout this paper we
use a fixed tilesize of \mbox{$32^3$} lattice nodes. This version has
been compiled with the \smpss{} 2.5 compiler driver. Initial
performance measurements of the tiled version using data copies (see
section \ref{sec:copies}) showed a penalty of roughly $40\%$ with
respect to the aliased version (depending on tile- and total problem
size) and was not considered further. The third version is a hybrid
\emph{MPI/OpenMP} version.  It uses the same tiled code base as the
\starss{} version, but does a \verb!omp parallel do! on the tiles
loop, which implies a synchronisation before applying boundary
conditions.

The benchmarks have been performed at HLRS on the cluster
\laki{}\cite{laki:web}. \laki{} is a cluster of roughly 700 nodes
interconnected through an Infiniband network. Each node is a two-way NUMA node
and has two Nehalem X5560 sockets, with 4 cores each, resulting in 8 cores per
node. Each node has access to 12GB of memory, however, non-uniformly across
sockets.

The MPI ranks were placed in such a way that 
communication to off-node destinations was the same for all
three versions. Communication with on-node partners was done
explicitly through library calls for the MPI version. In the OpenMP
version, data is shared and synchronization done through explicit
OpenMP directives, while the StarSs version is self-synchronised
through implicit data dependencies.

\subsection*{Single-node strong scaling}

\begin{figure}[!t]\centering
\includegraphics[width=0.98\columnwidth]{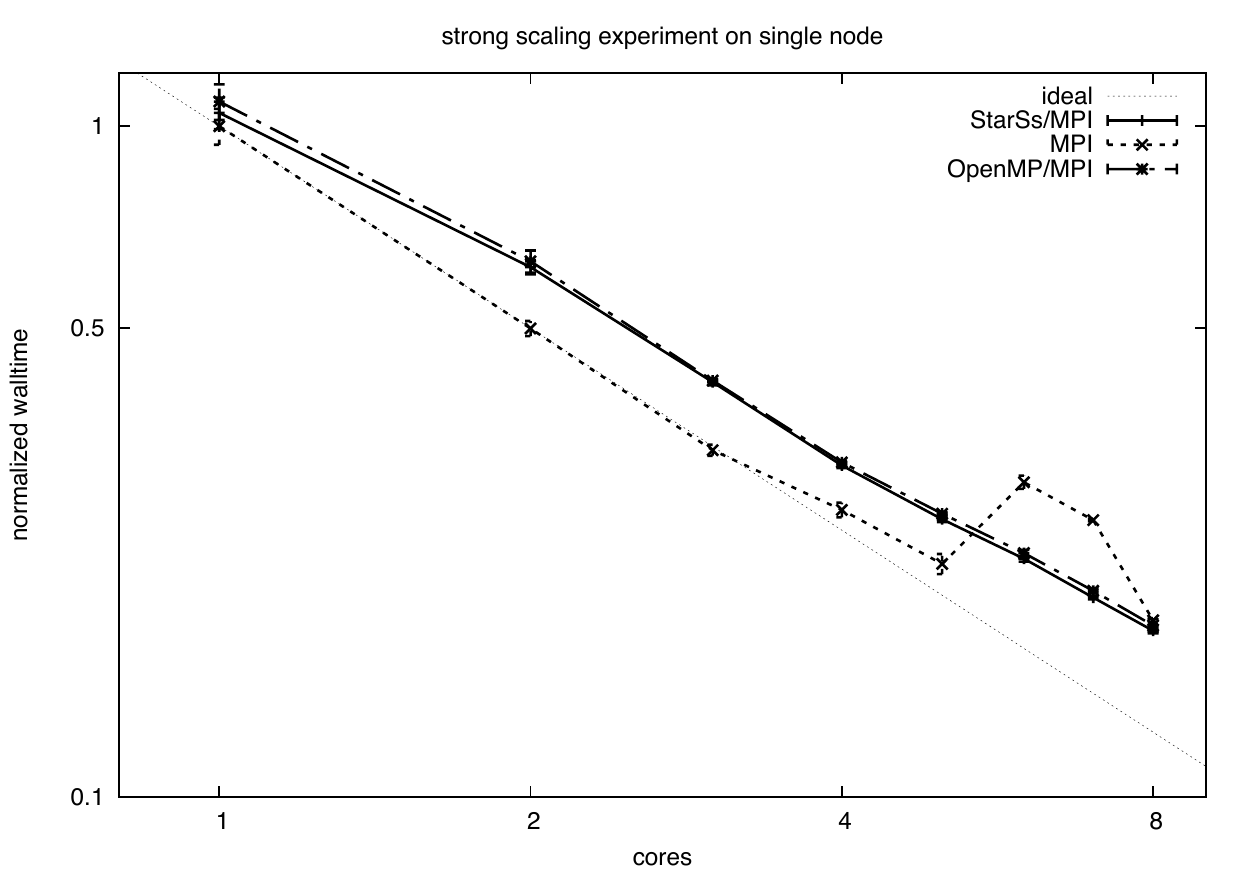}
\caption{Strong scaling experiment on a single node. Values are
  normalized to the execution time of the MPI-only version on a single
  core.}
  \label{fig:strong}
\end{figure}

\begin{figure}[!t]\centering
\includegraphics[width=0.98\columnwidth]{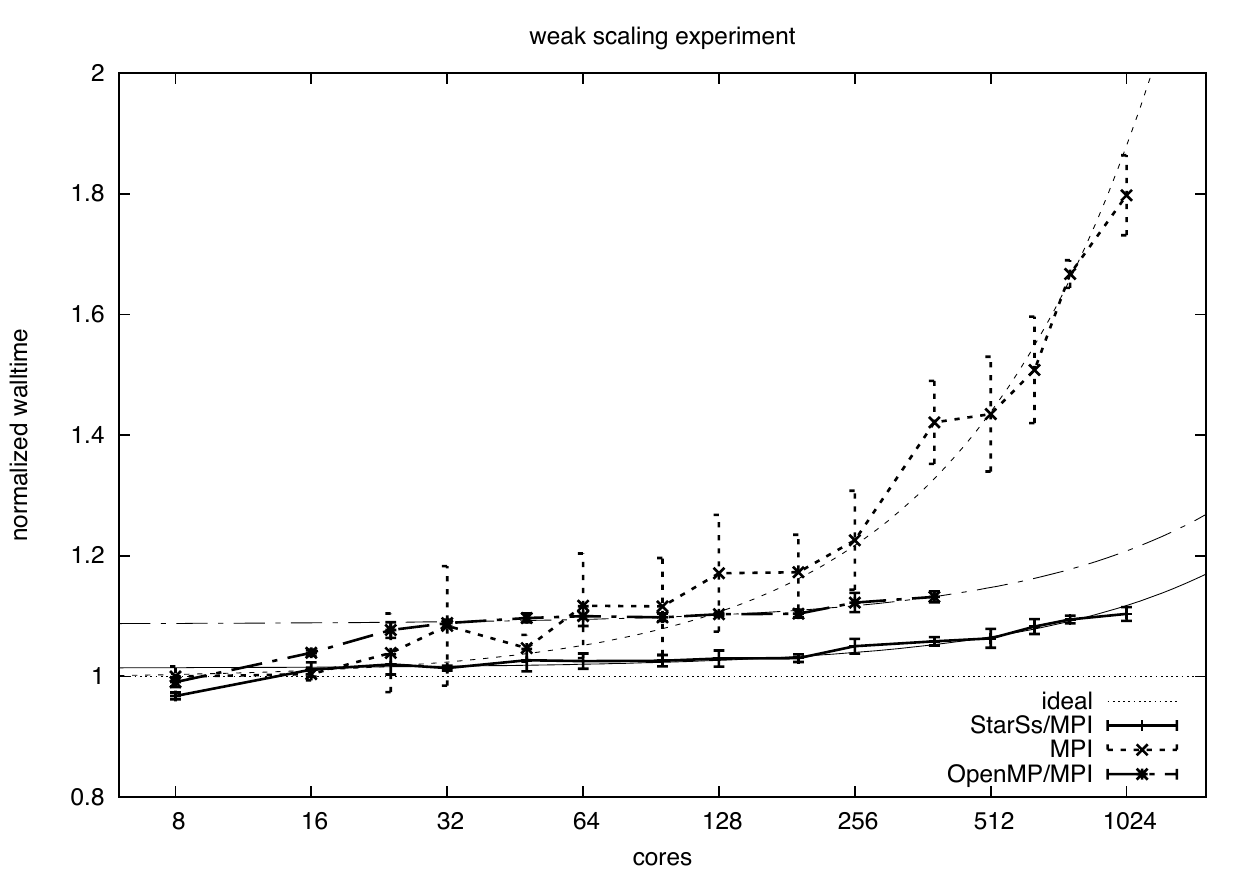}
  \caption{Weak scaling experiment on a cluster of Nehalem nodes
    plotted over number of cores. Values are normalized to the
    execution time of the MPI-only version on 8 cores.}
  \label{fig:weak_cores}
\end{figure}

\begin{figure}[!t]\centering
\includegraphics[width=0.98\columnwidth]{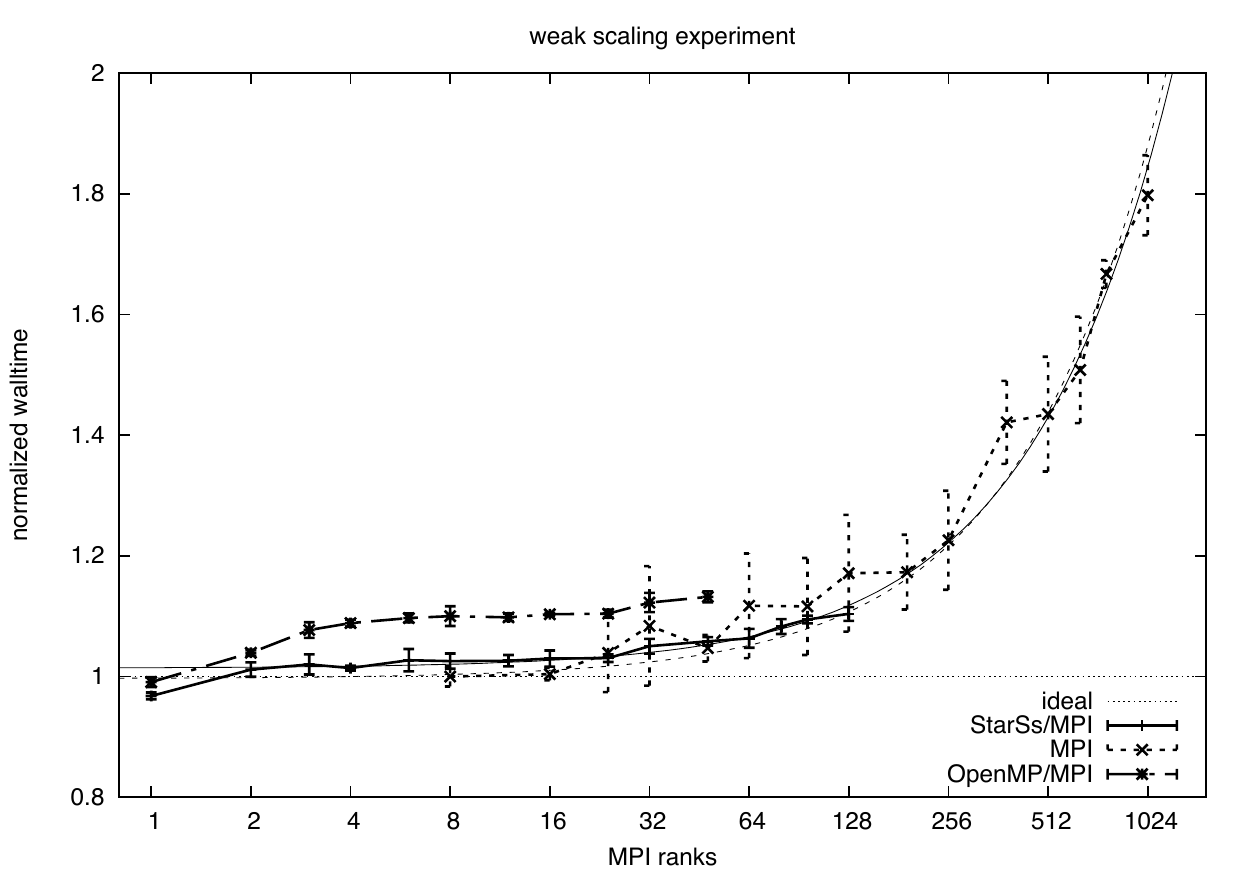}
  \caption{Weak scaling experiment on a cluster of Nehalem
    nodes plotted over number of MPI ranks (as opposed to cores in
    Fig. \ref{fig:weak_cores}). The same normalization as in Fig.~\ref{fig:weak_cores} is used.}
  \label{fig:weak_ranks}
\end{figure}

The first experiment we carried out was to measure the behaviour of the codes
under strong scaling on a single node.
The result for a \mbox{$256\times256\times224$} lattice is shown in
Fig. \ref{fig:strong}. This is the largest problem size that can be
run on a single Laki node. We have timed $50$ time-steps, including the
initialisation phase and a single phase of output at the very
end. Each measurement point is the mean of a large number of runs;
runs were done in several batches on different days in order to sample
a range of realistic load patterns of the cluster. The measurement
error is taken to be the statistical standard deviation. We have
normalised all measurements to the execution time, $t_1 = 115.04 s$,
of the MPI-only version running on a single MPI rank. In the
Lattice-Boltzmann community it is good practice to compare the performance of
different codes by the number of lattice-node updates per second
(MLUPs, mega lattice updates per second); our normalisation then
corresponds to $6.4$ MLUPs per core\footnote{These values are lower
  than those quoted in the literature for highly optimised LBM
  codes. Gains in scalar performance have the price of higher
  complexity of the source code which we wanted to avoid in this
  paper. Yet, all techniques discussed here can in principle be
  applied to highly optimized codes as well.}.

The first thing to notice from Fig. \ref{fig:strong} is that on a
single core there is no significant difference in the performance of
the three version, as the difference is smaller than the measurement
errors. Looking past measurement errors, one can see a performance
drop from MPI through MPI/\starss{} to MPI/OpenMP of less than $8\%$.

Doubling the number of cores from one to two, the MPI version scales
nearly perfectly, while both hybrid versions have significantly lower
scaling efficiency. Both hybrid versions, i.e. MPI/\starss{} and
MPI/OpenMP, scale with high efficiency beyond two cores up to the full
complement of $8$ cores, but remain below the performance of the MPI
version up to $6$ cores. The efficiency of the MPI versions initially
is very good, but then drops noticeably beyond $4$ cores. In fact,
there is only little overall speedup from $5$ to $8$ cores with an
intermediate increase in execution time.  When $8$ cores are reached,
both hybrid versions have made up on the efficiency they lost initially
compare to the MPI version and perform slightly better. This slight
performance advantage is not significant for the MPI/OpenMP
version. The performance advantage of the MPI/\starss{} version over
MPI-only is slight, but statistically significant.

Of the three versions, the MPI-only version displays the largest
scatter in execution times by a factor of at least three compared to
the hybrid versions. The measurement error of the hybrid \starss{}
version is relatively high for a single and two cores, but then falls
off noticeably.

The hybrid versions do not seem to be greatly affected by the NUMA
nature of the \laki{} nodes, nor the limited memory bandwidth, at
least not more than the MPI version. In fact, pinning MPI
processes to cores or sockets does not increase performance for the
MPI version and is difficult\footnote{in the sense of not easily
  portable across platforms} to implement particularly for the hybrid
versions.  We thus decided to use $8$ threads per MPI process for the
hybrid versions.

\subsection*{Multi-node weak scaling}
A weak scaling experiment with
\mbox{$32\times256\times224$} lattice nodes per core was done; core
counts range from $8$ (a single node) to $1024$ cores. Again we have
normalised all measurements to the execution time of the MPI-only
version, this time running with $8$ cores, i.e. $t_8 = 21.13 s$, which
corresponds to $4.3$ MLUS per core.

The results of the weak scaling experiment are presented in Figs.
\ref{fig:weak_cores} and \ref{fig:weak_ranks}.  Also in this batch of
runs, the hybrid MPI/\starss{} version tends to be slightly faster than
the MPI version, albeit only with a low significance. Increasing the
number of cores, the execution times of all three versions
can be modelled by smooth curves as discussed in the next section
\ref{sec:discuss}. No sudden change of scaling efficiency is observed
in any of the versions. Only the MPI/OpenMP version scales at
lower efficiency below $32$ cores, incurring a performance penalty of
$10\%$ compared to the MPI/\starss{} version. The scaling
efficiency of the hybrid MPI/\starss{} version is
significantly better than the one of the MPI-only version. At $1024$
cores the execution time of the MPI version has gone up by a factor of roughly
$1.8$ while the hybrid version is still running within a factor $1.1$
of the ideal execution time. So far, we have not been able to collect
sufficient measurements for the MPI/OpenMP version beyond $384$
cores. However, the data we have seems to indicate that also this
versions scales considerably better than the MPI-only version, but
due to the low scaling efficiency at small numbers of cores, it ends up
running at a factor of only $1.2$ of the ideal execution time.

As with the strong-scaling experiment, the scatter of execution times is
significantly larger for the MPI than for the hybrid versions; typically by a
factor of $5$ or higher.

\section{Discussion}
\label{sec:discuss}

\subsection{Multi-node scaling efficiency}
\label{sec:discuss_multi}
In this section we will discuss the experimental results presented in the
previous section in some more detail, particular in terms of scaling
efficiency. To quantify this, we start out by modelling the total execution
time $T(n)$ of the weak-scaling experiments by a parametric function in the
number of cores $n$ given as 
\begin{equation}
\label{eq:tweak}
  \frac{T^i(n)}{T_8} = \tau^i_0 + \mu^i \, n,
\end{equation}
where the index $i$ denotes the three versions (m, s, o for
MPI, MPI/\starss{}, and MPI/OpenMP, respectively); $T_8$ is the
normalization of the weak-scaling experiments (chosen to be the
execution time of the MPI-only version at 8 cores). The parameter
$\tau^i_0$ quantifies a constant contribution to the execution time, and
$\mu^i$ one that scales linearly in the number of cores. Fits of
the measurements to this parametric function are shown together with
the data in Fig. \ref{fig:weak_cores}.
Higher order functions with terms proportional to
$n\log n$, $n^2$, etc, did not improve any of the fits
significantly.

From the fit, the linear contribution to the execution time of the
MPI/\starss{} version is $\mu^s = (1.0 \pm 0.1) \times 10^{-4}$,
for the MPI-only version it is $\mu^m = (8.6 \pm 0.2) \times
10^{-4}$, for the MPI/OpenMP version it is $\mu^o = (1.2 \pm 0.2)
\times 10^{-4}$; values are rounded to the nearest digit. This
means, that the execution time of the weak-scaling experiment will
have doubled with respect to the reference (i.e. 8 cores) at ~$1160$
cores for the MPI version, at ~$8470$ cores for the hybrid MPI/OpenMP
version, and at $9860$ cores for MPI/\starss{}.

The ratio is $\mu^m/\mu^s = 8.5 \pm 0.5$, which is the
number of \starss{} threads running inside an MPI process for the hybrid
version. In other words, the hybrid version scales, in terms of MPI processes,
exactly the same way as the MPI-only version. However, it uses all of the
potential of the hybrid approach and scales a factor of 8 more efficiently in
terms of number of cores. This circumstance can be nicely illustrated by plotting
the performance data over the number of MPI ranks (see Fig.
\ref{fig:weak_ranks}) rather than over number of cores (see Fig.
\ref{fig:weak_cores}). In terms of MPI ranks, the MPI-only and the hybrid
MPI/\starss{} version scale nearly identically.
Compared to MPI/\starss{}, the MPI/OpenMP version
scales slightly less efficient with $\mu^m/\mu^o = 7.3 \pm 1.0$ yet not
significantly.

Under weak-scaling, the time a single core spends doing actual
computation is constant, i.e. independent of the total number of
cores. The total execution time may deviate from this ideal value due
to communication time. Our parametric function eq (\ref{eq:tweak}) can
thus be interpreted in terms of a constant computation time $\tau^i_0$
(including any constant-time communication contribution), plus a
communication term $\mu^i \, n$ which is proportional to the number of
cores. As the MPI domain decomposition is one-dimensional, the linear
communication time found in our scaling curves suggest, that in the
current MPI version of the code, the next-neighbour communication is
serialized across the communicator. And indeed, the original code
uses the same rank as source and destination argument for the
\verb!MPI_sendrecv()! call.  Obviously, this is simple to remedy and
would make the communication time a constant, independent of the number
of cores. Note, that this statement is true for all three versions of
the code as it affects only the communication between MPI ranks,
and does therefore not change the relative performance of the different
versions.

\subsection{Single-node performance}
\label{sec:discuss_single}

Next we would like to address the single-node performance of the
different code versions. As noted above, the performance of the three
versions on a single node is very similar. Assuming that measurement
errors were negligible, the \starss{} version is roughly $5\%$ faster
than the MPI version. These two versions differ in 1) the runtime
environment, and 2) the tiling in the code. Any of these two
could be causing the difference in performance.

We have therefore compiled the tiled code (as used for the \starss{}
version) with the same MPI wrapper as our orignal MPI-only
version. Given the measurement errors, the single-core performance of
this tiled MPI version is indistinguishable from the single-core
performance of the other three versions discussed in this paper. At
this level we cannot measure any statistically significant effects of
the respective runtimes, i.e. any overheads of SMPSs are negligible
(within the established measurement error estimates) for the problem
size and task granularity under consideration. 

At 8 cores, however, the hybrid MPI/\starss{} and the tiled MPI
version have a significant performance advantage over the original
non-tiled MPI code. While our data hints at a slight advantage of the
tiled MPI version also over the tiled hybrid version, the difference
is not significant. 

We have modelled the total execution time $T(n)$  of the strong
scaling experiment by a parametric function in the number of cores $n$
as
\begin{equation}
\label{eq:tstrong}
  \frac{T^i(n)}{T_1} = \frac{\tau^i_0}{n} + \beta^i + \eta^i \, (n-1),
\end{equation}
where the index i denotes any of the three versions (m, s, o for MPI,
MPI/StarSs, and MPI/OpenMP, respectively); $T_1$ is the normalization
of the weak-scaling experiments (chosen to be the execution time of
the MPI-only version at 1 core). The parameters $\tau^i_0$ relates the
execution time of a particular version to the reference, $\beta^i$
parametrizes constant-time contributions as for instance memory
bandwidth requirements that cannot be parallelized, and $\eta^i$
quantifies deviations from the ideal speedup that scales linearly in the
number of cores. This parametric function was used to fit the
measurements of the strong scaling experiment. Terms proportional to 
$n \log n$, $n^2$, etc, did not improve any of the fits
significantly.\footnote{Note, that for fitting the MPI version we have
  only taken into account core numbers equal to powers of two.}  In
all cases the parameter $\beta^i$ is not well constrained by the data
and consistent with $\beta^i=0$. We interpret the term $\eta^i \,
(n-1)$ as a measure for contention amongst the cores during accesses to
the memory system, either the cache or the main memory. The ratios
$\eta^m/\eta^s = 2.3 \pm 0.2$ and $\eta^m/\eta^o = 1.8 \pm 0.2$,
indicate that the hybrid versions scale significantly better on a
single node than the MPI-only version. If our interpretation of the
parameter $\eta^i$ is correct, we conclude that the MPI version
suffers stronger from memory contention.

Our application is memory bandwidth limited -- a
resource that decreases in proportion to the number of cores that need
to share it. The higher re-use of cached memory lines by the tiled
algorithm leads to a more efficient use of memory bandwidth.  We thus
argue that high-level tiling of the algorithm is beneficial from a
cache re-utilization perspective\footnote{Naturally, this high-level
  tiling of the algorithm may not substitute low-level loop-blocking
  tailored to specific cache hierarchies.} also for the MPI-only
version and not just an up-front investment with the sole purpose of
applying the task-based programming model \starss{}. In this view, the
porting effort of \starss{}, i.e. code annotations for task
identification and directionality of arguments, is minimal. 

Finally, we would like to address the peculiar performance measurements
of the MPI-only version between $5$ and $7$ cores. The drastic
increase in execution time stems from an imbalanced domain
decomposition; the domain (a power of 2) simply cannot be divided into
equal sub-domains for $5$, $6$, or $7$ cores. The same is true for $3$
cores also, but there the load-imbalance is relatively small.

\subsection{Overlapping computation and communication}
\label{sec:discuss_overlap}

One of the goals when optimizing MPI code is to hide the communication
latency by overlapping computation with communication. In our
application MPI calls can be issued as soon as the outermost lattice
nodes of the domain have been calculated. One possible implementation
is to split the low-level loops in the computation subroutines
to do the necessary iterations first, issue non-blocking MPI calls,
and only then do inner loops. In general this will lead to obfuscated,
bloated code, in particular if there are many computation subroutines.

An algorithm which is tiled on a high semantic level -- as ours -- suggests
a different approach. Since all calculations on a specific \verb!tile!
are done with a single subroutine call, it is sufficient to reorder a
single loop, namely the one over all tiles. One still has to use
non-blocking MPI and wait on the messages at a suitable point in the
code.   

In our hybrid MPI/\starss{} version we have followed a variation of
this scheme. We introduced two disjoint sentinels, representing
the outermost tiles and the remaining inner tiles respectively, 
and called the computation routine \verb!stream_collide()! with the
proper sentinel. The subroutines \verb!boundaries()! and
\verb!exchange_ghosts()! then carry only the sentinel representing the
outer tiles. In this way we arrive at a dependency tree, that allows to
issue boundary conditions and MPI communication as soon as
computations on the outer tiles are done; inner tiles are calculated in
parallel. No modifications to the original blocking MPI calls are
necessary. 

The amount of communication time that can be hidden depends
on the ratio of inner to outer tiles and the number of cores
working on them. Specifically, we can hide roughly $250\, \mathrm{ms}$
communication time per timestep. As the communication time increases
with the number of cores, this limit is reached between $128$ and
$256$ cores. Below that, communication is completely overlapped with
calculations. With this communication scheme, fitting the parametric function
eq. \ref{eq:tweak} to data points between $32$ and $256$ cores,
results in a much flatter curve for the MPI/\starss{} version; with
$\mu´^s \sim 3 \times 10^{-5}$ the total execution time would double
only at beyond $32000$ cores based on an extrapolation of the curve. This makes
the hybrid MPI/\starss{} version scale much better than the MPI-only
version in this range. 

\section{Conclusions}

We set out to investigate if the hybrid MPI/\starss{} programming
model allows typical applications to scale to larger number of
cores than MPI-only applications and how much porting effort this
model requires.
Most research groups applying computational methods to
solve their domain specific problems will have only limited resources
to do such a port.

In our LBM example we have shown that the hybrid version,
combining MPI with the task-based programming model \starss{}, is a)
competitive with the MPI-only version on a single node, in fact
outperforms it, b) leverages the full scaling efficiency
across nodes, and c) results in much more stable execution times
reducing the effect of load-imbalance (internal or external as,
e.g., due to OS jitter). 
We have also evaluated a hybrid MPI/OpenMP version based on the same
code refactoring done for MPI/\starss{}.

Both hybrid versions are competitive with the MPI-only version on a
single core regarding strong scaling. The overhead of \starss{} task
instantiation, management, and scheduling is negligible for
the chosen task granularity. In fact, the \starss{} version is
slightly faster than the MPI version when the full node is used. This
stems from a better cache re-use due to tiling which leads to 
a lower contention among cores on the memory system. The same is
true for the hybrid MPI/OpenMP version but to a somewhat lower degree.

Doing hybrid distributed/shared memory parallelisation is often
motivated by a reduction of MPI ranks involved in a calculation in
order to reduce communication times which depend on the number of
ranks involved. At the very least, hybrid models should allow to scale
a given application to a larger number of cores than the MPI-only
version. Ideally, the range to which an application scales is extended
by a factor equal to the number of threads running inside a single MPI
rank.  Scalar overheads of the hybrid model should
remain small in order not to counteract the higher scaling efficiency
-- doing so increases the parallel efficiency of the application
to a much larger number of total {\em cores}, in particular on
multi- or many-core systems.

We have shown that the hybrid versions, in particular the
MPI/\starss{}, make up on their promise and leverage the full potential
by allowing to scale out by a factor equal to the number of cores on a
single node. In addition, \starss{} allows to hide communication
latencies in a straightforward way. In our case this turned out not to be very
relevant, as the communication time increases very rapidly and
eventually can not be hidden. Up to this stage, however, the weak
scaling curve is practically flat.  At scale, it is more efficient to
use the hybrid version than the MPI-only version.
 
In a future paper we will investigate different methods to overlap
communication and computation in more detail. In \starss{} this is
particularly attractive as the dynamic scheduling driven by data
dependencies allows to hide communication latencies as long as
sufficient concurrency is available. Also, the \smpss{} runtime
specifically allows to off-load MPI communication onto a special
communication thread \cite{MLAV10} that, in most cases, will not use
cycles and allows to have multiple transfers on the fly at the same
time. In that case the programmer does not need to bother with
non-blocking asynchronous MPI calls, but may continue to use the most
simple MPI calls and still get asynchronous communication with latency
hiding.

The MPI version has a much larger scatter of execution
times for the weak and the strong scaling experiments than both of the
hybrid versions. While we do not have execution traces to support our
hypothesis, we argue that the scatter stems from load imbalances due
to OS jitter. All code versions compete for resources with the OS,
however, any delay caused to a single task of the hybrid versions will
not affect the tasks running on the remaining cores, yet for the MPI
version all cores have to idle eventually waiting on the delayed core
to finish its work thus multiplying the initially small disturbance.   
In any case hybrid models help with application specific load
imbalances.

Finally, the porting effort was limited to tiling the algorithm, which
is not particularly difficult for LBM codes, especially in Fortran
where aliasing of sub-array can be used. Given the fact, that the
MPI-only version also benefits from this tiling, it is very much worth the
effort. Once done, executing the code in hybrid mode, either in
combination with OpenMP or \starss{} allows to scale out to much
larger number of cores. A future work will investigate if and how
\starss{} can be used to do temporal blocking in addition to the spatial
blocking presented in this work.

\section{Acknowledgments. }
\begin{sloppypar}
This work was supported by the European Community's Seventh Framework
Programme [FP7-INFRASTRUCTURES-2010-2] project TEXT under grant
agreement number 261580 and by the project PRACE-1IP under contract RI-261557.
\end{sloppypar}

\bibliographystyle{IEEEtran}
\bibliography{library}

\begin{thebibliography}{10}
\providecommand{\url}[1]{#1}
\csname url@samestyle\endcsname
\providecommand{\newblock}{\relax}
\providecommand{\bibinfo}[2]{#2}
\providecommand{\BIBentrySTDinterwordspacing}{\spaceskip=0pt\relax}
\providecommand{\BIBentryALTinterwordstretchfactor}{4}
\providecommand{\BIBentryALTinterwordspacing}{\spaceskip=\fontdimen2\font plus
\BIBentryALTinterwordstretchfactor\fontdimen3\font minus
  \fontdimen4\font\relax}
\providecommand{\BIBforeignlanguage}[2]{{%
\expandafter\ifx\csname l@#1\endcsname\relax
\typeout{** WARNING: IEEEtran.bst: No hyphenation pattern has been}%
\typeout{** loaded for the language `#1'. Using the pattern for}%
\typeout{** the default language instead.}%
\else
\language=\csname l@#1\endcsname
\fi
#2}}
\providecommand{\BIBdecl}{\relax}
\BIBdecl

\bibitem{B+09}
P.~Balaji, D.~Buntinas, D.~Goodell, W.~Gropp, S.~Kumar, E.~Lusk, R.~Thakur, and
  J.~L. Tr\"{a}ff, ``Mpi on a million processors,'' in \emph{Proceedings of the
  16th European PVM/MPI Users' Group Meeting on Recent Advances in Parallel
  Virtual Machine and Message Passing Interface}.\hskip 1em plus 0.5em minus
  0.4em\relax Berlin, Heidelberg: Springer-Verlag, 2009, pp. 20--30.

\bibitem{*Ss:BPBL06}
P.~Bellens, J.~M. Perez, R.~M. Badia, and J.~Labarta, ``Cellss: a programming
  model for the cell be architecture,'' in \emph{ACM/IEEE CONFERENCE ON
  SUPERCOMPUTING}.\hskip 1em plus 0.5em minus 0.4em\relax ACM, 2006, p.~86.

\bibitem{HKR11}
M.~Hasert, H.~Klimach, and S.~Roller, ``Caf versus mpi - applicability of
  coarray fortran to a flow solver,'' in \emph{Lect. Notes. Comput. Sci.}\hskip
  1em plus 0.5em minus 0.4em\relax Springer, 2011.

\bibitem{BGK54}
P.~L. Bhatnagar, E.~P. Gross, and M.~Krook, ``{A Model for Collision Processes
  in Gases. I. Small Amplitude Processes in Charged and Neutral One-Component
  Systems},'' \emph{Phys. Rev.}, vol.~94, no.~3, pp. 511--525, 1954.

\bibitem{BNGK11a}
S.~Brinkmann, C.~Niethammer, J.~Gracia, and R.~Keller, ``Temanejo - a debugger
  for task based parallel programming models,'' in \emph{ParCo2011: Proceeding
  of the International Conference on Parallel Computing}, 2011, accepted.

\bibitem{debugger:web}
\BIBentryALTinterwordspacing
``\textsc{Temanejo} -- a starss debugger,'' Jul. 2011. [Online]. Available:
  \url{http://www.hlrs.de/organization/av/amt/projects/text/}
\BIBentrySTDinterwordspacing

\bibitem{*Ss:SMPSs25}
\BIBentryALTinterwordspacing
(2011, Jul.) Smpss distribution v2.5. [Online]. Available:
  \url{http://www.project-text.eu/sites/default/files/SMPSs-2.5.tar.gz}
\BIBentrySTDinterwordspacing

\bibitem{*Ss:PBL08}
J.~Perez, R.~Badia, and J.~Labarta, ``A dependency-aware task-based programming
  environment for multi-core architectures,'' in \emph{Cluster Computing, 2008
  IEEE International Conference on}, 29 2008-oct. 1 2008, pp. 142 --151.

\bibitem{laki:web}
\BIBentryALTinterwordspacing
``Nec nehalem cluster,'' Jul. 2011. [Online]. Available:
  \url{http://www.hlrs.de/systems/platforms/nec-nehalem-cluster/}
\BIBentrySTDinterwordspacing

\bibitem{MLAV10}
\BIBentryALTinterwordspacing
V.~Marjanovi\'{c}, J.~Labarta, E.~Ayguad\'{e}, and M.~Valero, ``Overlapping
  communication and computation by using a hybrid mpi/smpss approach,'' in
  \emph{Proceedings of the 24th ACM International Conference on
  Supercomputing}, ser. ICS '10.\hskip 1em plus 0.5em minus 0.4em\relax New
  York, NY, USA: ACM, 2010, pp. 5--16. [Online]. Available:
  \url{http://doi.acm.org/10.1145/1810085.1810091}
\BIBentrySTDinterwordspacing

\end{thebibliography}

\end{document}